\documentstyle[amssymb,prb,aps,preprint]{revtex}

\begin{document}
\title{Low-temperature spin relaxation in n-type GaAs}
\author{R.I.Dzhioev,$^{1}$ K.V.Kavokin,$^{1,2}$ V.L.Korenev,$^{1}$ M.V.Lazarev,$^{1}$
B.Ya.Meltser,$^{1}$ M.N.Stepanova,$^{1}$ B.P.Zakharchenya,$^{1}$ D.Gammon,$%
^{3}$ D.S.Katzer$^{3}$}
\address{$^{1}${\it A.F.Ioffe Physico-Technical Institute, 194021 Politechnicheskaya }%
26, \\
St. Petersburg, Russia}
\address{$^{2}$ School of physics, University of Exeter, Stocker Road, Exeter, EX4\\
4QL, UK}
\address{$^{3}${\it Naval Research Laboratory, Washington DC, USA }}
\date{Received {\today}}
\maketitle

\begin{abstract}
Low-temperature electron spin relaxation is studied by the optical
orientation method in bulk n-GaAs with donor concentrations from 10$^{14}$cm$%
^{-3}$ to 5$\cdot $10$^{17}$ cm$^{-3}$. A peculiarity related to the
metal-to-insulator transition (MIT) is observed in the dependence of the
spin lifetime on doping near n$_{D}$ = 2$\cdot $10$^{16}$cm$^{-3}$. In the
metallic phase, spin relaxation is governed by the Dyakonov-Perel mechanism,
while in the insulator phase it is due to anisotropic exchange interaction
and hyperfine interaction
\end{abstract}

\narrowtext

\section{Introduction}

The research on the physics of non--equilibrium spin in semiconductors has
been conducted for more than 30 years, since first experiments on optical
orientation of electron and nuclear spins, performed by G.Lampel \cite{1} in
Si, by R.Parsons \cite{2} in GaSb, and by Zakharchenya et al \cite{3} in
GaAs. Basic facts and a considerable body of experimental and theoretical
results related to bulk Group III-V semiconductors are collected in the
monograph ''Optical Orientation'' \cite{4} issued in the 80-th. Later on,
much new information concerning mainly low-dimensional structures has been
obtained. Nevertheless, there remain gaps in this knowledge, that have
become visible with the emerging of an application- directed angle on
spin-related phenomena (spintronics) \cite{5}. Though it was known to
specialists that n-type semiconductors demonstrate, generally, extended spin
lifetimes \cite{6,7}, a recent finding of over- 100ns spin memory\cite{8} in
bulk gallium arsenide with the donor concentration of 10$^{16}$cm$^{-3}$
became a surprise, and attracted an increased attention to n-type
semiconductors as a possible base for spintronic devices. It was suggested
that the spin lifetime as a function of donor concentration $n_{D}$ has a
maximum at $n_{D}$ near 10$^{16}$ cm$^{-3}$. Later on, even longer spin
lifetime of nearly 300ns was reported in a GaAs/AlGaAs heterostructure \cite
{9,10}. However no detailed experimental or theoretical study of the
dependence of the electron spin relaxation on doping has been done so far.
This paper is aimed at filling this gap. The choice of GaAs for this study
is justified not only by its prospective spintronic applications, but also
by the fact that the physics of spin systems in this semiconductor is
otherwise very well studied. Once an understanding of the spin relaxation
processes is reached for GaAs, it can be easily extended to other
semiconductors. We use optical orientation technique to measure the
concentration dependence of the electron spin relaxation time in n-type
epitaxial layers of GaAs at liquid-helium temperatures. Comparison of the
experimental data with theory reveals the main mechanisms of spin relaxation
relevant in this temperature range, and determines the limits to the spin
lifetime in bulk n-type semiconductors.

\section{Samples and experimental technique}

We used 2mkm thick layers of GaAs between AlGaAs barriers, grown by the
molecular-beam epitaxy ($n_{D}=5.5\times 10^{14},2\times 10^{16},4.6\times
10^{16},5.6\times 10^{16},9\times 10^{16}cm^{-3}$); 20 mkm thick layers
grown by liquid-phase epitaxy ($n_{D}=1\times 10^{15},2\times
10^{15},2.1\times 10^{15},4\times 10^{15},4.5\times 10^{15},7\times
10^{15},1\times 10^{16},1.6\times 10^{16},2.8\times 10^{16}$); a bulk
Chochralskii-grown crystal ($n_{D}=5\times 10^{17}cm^{-3}$); a 0.1mkm thick
GaAs buffer layer of a multi-quantum well structure ($n_{D}=1\times
10^{14}cm^{-3}$).

The samples were placed in a liquid-helium cryostat and pumped by a tunable
Ti- sapphire laser, with the circular polarization of light being alternated
in sign at a frequency of 26.61 kHz with a photoelastic quartz modulator.
This allowed us to eliminate the effect of the lattice nuclear polarization
on the optical orientation of the electrons (Chap.5 and 9 of Ref.4). The
geomagnetic field was compensated to a level of not over 0.1 G at the
sample. The PL polarization was measured in the reflection geometry by a
circular-polarization analyzer. The PL was dispersed by a double-grating
spectrometer (5 $\AA /mm$). A two-channel photon counting device
synchronized with the quartz modulator provided measurement of the effective
degree of circular polarization $\rho _{c}=\frac{I_{+}-I_{-}}{I_{+}+I_{-}}$
, where $I_{+}$ and $I_{-}$ are the intensities of the $\sigma _{+}$ PL
component under the $\sigma _{+}$ and $\sigma _{-}$ pumping, respectively. $%
\rho _{c}$ may be considered as a Stokes parameter characterizing the PL
circular polarization. It is proportional to the amplitude value of the
average electron spin induced by the alternate-polarized pump light.

The method of determination of the spin relaxation time in n-type
semiconductors by steady-state optical orientation is based on the following
physical grounds \cite{4,9}. After creation of an electron-hole pair by
circularly polarized light, the hole rapidly loses the memory about its
initial spin state. Then it recombines with an electron, besides under low
pump intensity the probability of recombination with a photoexcited electron
is negligible as compared with the probability to recombine with one of the
unpolarized equilibrium electrons. Thus, spin- polarized photoexcited
electrons eventually substitute unpolarized equilibrium electrons, and spin
polarization accumulates in the crystal. If the density of photoexcited
carriers is spatially uniform, then, under cw excitation, the spin lifetime
is given by the expression:

\begin{equation}
T_{S}=\left( \frac{1}{\tau _{s}}+\frac{1}{\tau _{J}}\right) ^{-1}  \label{1}
\end{equation}

where $\tau _{s}$ is the spin relaxation time, $\tau _{J}=n/G$, $n$ is the
concentration of equilibrium electrons, and $G$ is the excitation density
(the rate of creation of photocarriers per unit volume). The suppression of
the electron spin orientation in transversal magnetic field (the Hanle
effect) in this simplest case is described by the Lorentz curve:

\begin{equation}
s_{z}(B)=s_{z}(0)\frac{1}{1+\left( \mu _{B}gBT_{S}/\hbar \right) ^{2}}
\label{2}
\end{equation}

where $B$ is the magnetic field, $\mu _{B}$ is the Bohr magneton, and $g$ is
the electron g-factor.

If the concentration of photoexcited carriers significantly changes over the
region where electron spins are polarized, it is not possible to describe
the entire ensemble of electrons by the unique $\tau _{J}$. In this case,
the Hanle curve is no longer Lorentzian. Also, spin diffusion may result in
non-Lorentzian Hanle curves on the high-energy side of the PL spectrum \cite
{7}. However, in our experiments none of these effects have been observed:
within the experimental accuracy, the Hanle curves were Lorentzian and
identical within the width of the PL lines.

As in GaAs the g-factor is known, Eq.(2) allows to determine $T_{S\text{ }}$%
from the Hanle effect: the half-width of the curve, $B_{1/2}=\frac{\hbar }{%
\mu _{B}g}T_{S}^{-1},$is proportional to the inverse spin lifetime. It
follows from Eq.(1) that $T_{S}$ and, therefore, the width of the Hanle
curve, depends on the excitation intensity. To obtain the value of $\tau _{s}$
, one should take $T_{S}$ in the low-pump limit.

\section{Experimental results and discussion}

PL spectra (Fig.1a) of samples with low doping level ($n_{D}$ 
\mbox{$<$}%
10$^{15}$ cm$^{-3}$) consist of up to 3 overlapping lines corresponding, to
the best of our knowledge \cite{11}, to recombination of free excitons ($X$, 
$1.5155eV$), of excitons bound to neutral donors ($D^{0}X,1.5145eV$), and of
excitons bound to charged donors ($D^{+}X,1.5136eV$). The free exciton
recombination forms the high-energy wing of the spectrum. With the increase
of donor concentration, these lines merge into one broad line. Under optical
orientation conditions, PL is circularly polarized. Both the zero-field
polarization and the width of the Hanle curve decrease with the decrease of
excitation intensity, which is typical for optical orientation of
equilibrium electrons \cite{4}. In samples with low donor concentration, the
polarization degree is the highest at the high- energy wing corresponding to
excitonic transitions, then it falls down to zero at the maximum of $D^{0}X$
line, and slightly increases with further decrease of the PL photon energy.
The dip in the spectral dependence of $\rho _{c}$ results from coupling of
spins of the two electrons in the $D^{0}X$ complex into a singlet state.

In heavily-doped samples, the circular polarization is only observed at the
high-energy wing of the spectrum (Fig.1b). This behavior reflects the
Fermi-statistics of delocalized electrons in degenerate semiconductor
crystals: only Fermi-edge electrons may have a non-zero average spin. The
dependence of the polarization degree on the transversal magnetic field (the
Hanle effect) is the same for all the PL energies. This is an evidence that,
under sufficiently low excitation densities we used, the PL polarization at
all the photon energies reflected the state of the same spin reservoir,
namely that of equilibrium electrons \cite{12}, and the differences in the
polarization degree were due to specific recombination conditions rather
than to spin dynamics. Respectively, measuring $T_{S}$ at the limit of low
pump density yielded the value of $\tau _{s}$ characterizing the electron
ensemble of the sample under study.

An example of the dependence of $T_{S}$ on pump intensity is shown in Fig.2.
The Hanle curve becomes steadily narrower with decreasing the intensity. The
half-width of the Hanle curve vs pump is plotted in the inset. It is well
fitted by a linear dependence, whose cutoff at zero pump gives the desirable
spin relaxation rate. This procedure was used to determine $\tau _{s}$ for
each of our samples. The results are shown in Fig.3. To fully represent the
available experimental information, we plot here also data from Ref.8
obtained by use of time-resolved pump-probe technique. In spite of a
considerable scattering of experimental points (this results, in our
opinion, mainly from errors in determination of the donor concentration, and
from incontrollable impurities present in the samples), they give an
unambiguous picture of spin relaxation over a wide range of doping. The most
remarkable feature of the concentration dependence of $\tau _{s}$ is that it
has {\it two} maxima. With the increase of doping from 10$^{14}$ cm$^{-3}$
upwards $\tau _{s}$ , being initially about $5\,ns$, becomes longer,
reaching values around $180\,ns$ at $n_{D}\approx 3\times 10^{15}$ $cm^{-3}$
, then decreases down to approximately $50$ $ns$ at $n_{D}\approx 1.5\times
10^{16}$ $cm^{-3}$. Further increase of the donor concentration results in
an abrupt three-fold rise of the spin relaxation time, followed by its
steady and steep decrease ($\tau _{s}$ becomes shorter by nearly four
decimal orders over the next two orders in the donor concentration). The
spin relaxation time is virtually the same at $2\,$and $4.2\,K$ , which
suggests that in this temperature range scattering by phonons has
practically no impact on the electron spin, and that, in heavily doped
samples, we observe spin dynamics of electrons obeying a degenerate
statistics .

We interpret this unusual concentration dependence as a manifestation of
three mechanisms of spin relaxation relevant for equilibrium electrons at
low temperature: hyperfine interaction with spins of lattice nuclei \cite
{13,14}, anisotropic exchange interaction of donor-bound electrons \cite{15}%
, and the Dyakonov-Perel mechanism \cite{16}. The maximum at $n_{D}=3\cdot
10^{15}cm^{-3}$ is due to a crossover between relaxation mechanisms
originating from the hyperfine interaction with lattice nuclei and from the
spin-orbit interaction. The peculiarity at $n_{D}=2\cdot 10^{16}cm^{-3}$ is
associated with the metal-to-insulator transition (MIT)\cite{EfrSh}. It
reflects the change of the specific mechanism through which the spin-orbit
coupling affects the spin lifetime: in the metallic phase it is the DP
mechanism, while in the insulator phase ($n_{D}<2\cdot 10^{16}cm^{-3}$) it
is the anisotropic exchange.

All the three mechanisms can be interpreted in terms of effective magnetic
fields acting upon the electron spin. Spin-orbit interaction in crystals
without inversion symmetry, like GaAs, is known to produce effective fields
determined by the direction and value of the electron wave vector ${\bf k}$.
Scattering by defects or phonons results in this field's rapid changing in
time; the spin is therefore exposed to a stochastic field which causes its
relaxation \cite{16}. This is referred to as the Dyakonov-Perel mechanism.
It has been shown that an analogous field affects the spin of an electron
tunneling through a potential barrier \cite{15}. As a result, the exchange
interaction of donor-bound electrons in GaAs turns out to be anisotropic,
and the flip-flop transition of spins of two electrons coupled by the
exchange interaction goes along with rotation of each of the spins through
the same small angle $\gamma \approx 0.01$, but in opposite directions. The
axis of the rotation, as well as the value of $\gamma $, depends on the
orientation of the pair of donors in the crystal. In the ensemble of
randomly distributed donors, this process leads to relaxation of the total
spin of the donor-bound electrons \cite{15}. Another contribution into the
spin relaxation rate of localized electrons comes from their interaction
with nuclear spins. As the donor-bound electron interacts with a great
number of nuclei, $N\approx 10^{5}$, the effect of nuclei upon the electron
spin ${\bf S}$ can be always presented as a Larmor precession of ${\bf S}$
in an effective ''hyperfine'' magnetic field with contribution of all the
nuclear spins within the electron orbit (Chapter 2 of Ref.4; Ref.13). The
hyperfine field produced by the mean-squared fluctuation of the nuclear spin
is equivalent to the combined action of $\sqrt{N}\approx 300$ spins, which
amounts to approximately 54 $Oe$ for GaAs \cite{10}.

One can see that these three mechanisms give the qualitative picture of the
concentration dependence of $\tau _{s}$, which is consistent with our
experimental observations. Indeed, at low donor concentrations electrons are
effectively isolated, and their spins precess independently in random static
nuclear fields. This results in disappearance of the most part of the
electron spin orientation within a few nanoseconds \cite{10,14}. Then, with
increasing donor concentration, electron wave functions begin to overlap,
and the isotropic exchange interaction brings about flip-flop transitions,
which results in dynamical averaging of the hyperfine interaction: the
electron spin ceases to be bound to a single donor and interacts with a
greater number of nuclei, so that the effect of nuclear-spin fluctuations
becomes smaller. As a result, $\tau _{s}$ increases. On the other hand,
stronger overlap of wave functions is accompanied by a greater probability
to lose spin orientation due to the anisotropic exchange interaction.
Eventually, the anisotropic exchange becomes stronger than the hyperfine
interaction, and the rise of the spin lifetime is changed for the decrease.
Finally, above MIT, the Dyakonov-Perel (DP) mechanism governs spin
relaxation. The increase of the Fermi wave vector with the electron
concentration makes the DP spin relaxation faster, and $\tau _{s}$ gets
steadily shorter. The discontinuity in the concentration dependence of $\tau
_{s}$, observed at MIT, suggests that at this concentration spin relaxation
in the insulator phase (via anisotropic exchange) is faster than in the
metallic phase (DP). This conclusion agrees with the results of theoretical
calculations for dielectric and metallic phases (see below); however, we
cannot propose any quantitative theory of spin relaxation in the MIT region.

A common feature of all the spin relaxation mechanisms based on spin
precession in random magnetic fields is that they can be suppressed by
applying a longitudinal magnetic field. Indeed, this is equivalent to
placing the electrons in a rotating frame, where transverse components of
random fields are reduced as a result of dynamical averaging. The
characteristic magnetic field required to suppress spin relaxation can be
found from the relation $\Omega _{L}\tau _{c}=1$ where $\Omega _{L}$ is the
Larmor frequency, and $\tau _{c}$ is the correlation time of the random
field. We performed experiments in longitudinal magnetic fields, placing our
samples into a superconducting solenoid immersed in liquid helium under
exhaust pumping (at 2K). This setup did not allow to measure the Hanle
effect; however we were able to detect changes in spin relaxation time by
measuring the dependence of $\rho _{c}$ on the magnetic field. Since we used
excitation with light of alternating helicity, and detected the polarization
signal at the modulation frequency (26.6 kHz), the field-induced circular
polarization of PL \cite{17} did not contribute into the measured signal,
which was, respectively, entirely due to optical orientation of electron
spins. The detected increase of $\rho _{c}$ with magnetic field was
therefore associated with suppression of spin relaxation, and characteristic
magnetic fields determined for each sample were used to calculate $\tau _{c}$%
. The results are shown by triangles in Fig.3. We were unable to measure $%
\tau _{c}$ for samples with donor concentration higher than 4$\cdot 10^{15}$
cm$^{-3}$ because strong magnetic fields required caused shifts of the PL
spectral lines, which resulted in strong parasite signals due to the
spectral dependence of $\rho _{c}$. Such measurements at higher donor
concentrations can be possibly done using time-resolved techniques. The
value of $\tau _{c}$ for the sample with donor concentration of $10^{14}$ cm$%
^{-3}$, where $\tau _{c}>\tau _{s}$, and the regime of isolated donors is
supposed to be realized \cite{10}, was calculated from experimental data by
use of a more complicated procedure, as described in details in Ref.10.

One can see that the measured values of $\tau _{c}$ fall into the nanosecond
and sub-nanosecond range. Therefore, $\tau _{c}$ cannot be associated with
the nuclear spin system which has much longer relaxation times (Chapter 2 of
Ref.4), and must be attributed to electrons. This means that $\tau _{c}$ is
in fact the local spin lifetime at a fixed donor; formally, this can be
written as a decay time of the electron-spin correlation function:

\begin{equation}
\tau _{c}=\frac{1}{{S\left( S+1\right) }N_{D}}\sum\limits_{i}\int%
\limits_{0}^{\infty }\left\langle {\bf S}_{i}(0){\bf \cdot S}%
_{i}(t)\right\rangle dt  \label{3}
\end{equation}

where angular brackets denote quantum-mechanical averaging, $i$ numerates
donors, $N_{D}$ stands for the total number of donors in the crystal.

Due to various spin-conserving processes providing spin transfer within the
impurity band, $\tau _{c}$ indeed can be much shorter than the spin lifetime
of the entire electron ensemble. For donor-bound electrons at low
temperature, the most relevant mechanism of spin transfer is exchange
interaction of electrons localized at adjacent donors. This conclusion is
qualitatively consistent with the steep decrease of $\tau _{c}$ with donor
concentration - this is a consequence of increased overlap of electron wave
functions. The estimation we performed using this model (see dotted line in
Fig.3; details of calculations are given in the following section), indeed
shows a good agreement with all the available experimental data on bulk
samples, i.e. at concentrations from $5.5\cdot 10^{14}cm^{-3}$ to $4\cdot
10^{15}cm^{-3}$. At lower $n_{D}$, the exponential concentration dependence,
characteristic for the exchange mechanism, gives very long $\tau _{c}$,
which becomes much longer than corresponding spin relaxation times at
concentrations of order and below $10^{14}cm^{-3}$. This fact suggests that
additional mechanisms of correlation decay may be significant at low donor
concentrations, where the exchange interaction is less effective. This
conclusion is backed by the data of Weisbuch\cite{6}, who reported the spin
relaxation time as long as 20ns in a bulk GaAs sample with $%
n_{D}=10^{13}cm^{-3}$. At such a low donor concentration, the regime of
isolated donors must have been realized, which would have resulted in a
shorter $\tau _{s}$, about 5ns, due to spin precession in the fluctuation
nuclear field\cite{10,14}. A longer time observed indicates that, most
likely, $\tau _{c}$ in that sample was rather short; however the specific
reason for shortening the correlation time is not clear. One of the possible
mechanisms, namely exchange interaction with free conduction-band electrons,
was studied in Ref.10. It was shown that additional electrons present in
space-charge layers of doped heterostructures can significantly reduce $\tau
_{c}$. In presence of additional electrons, the spin lifetime in a GaAs
layers in a MBE-grown multilayer structure (with the nominal doping level of 
$10^{14}cm^{-3}$) was as long as 290 ns, which corresponds to $\tau
_{c}\approx 0.1ns$. Recharging the GaAs layer under illumination allowed to
reduce the spin lifetime nearly 100-fold \cite{10}, down to 5ns, while $\tau
_{c}$ became as long as 17ns (these data are shown in Fig.3). The
correlation time of 17ns is still much shorter that what can be expected of
exchange interaction at $n_{D}=10^{13}cm^{-3}$. Possibly, some background
concentration of free electrons remained in the layer even under
illumination, which would have explained why $\tau _{c}$ was shorter than
expected in this specific sample. However, it remains unclear whether or not
delocalized electrons can be present in bulk samples at liquid-helium
temperatures. Our data do not give an unambiguous answer to this question,
and the issue of mechanisms of correlation decay in samples with low donor
concentrations remains open for future research.

\section{Theory}

\subsection{Isolating phase $(n_{D}<2\times 10^{16})$}

In order to estimate whether or not the exchange interaction can provide the
observed values of $\tau _{c}$, it is worth to note that the exponential
dependence of the exchange constant $J$ on the inter-donor distance must
result in an exponential decrease of $\tau _{c}$ with increasing donor
concentration. In the limit of extremely low concentrations, only nearest
neighbours contribute into the exchange interaction. The distribution
function of the distance to the nearest neighbour has the maximum at $%
r_{1}\approx 0.54n_{D}^{-1/3}$. At higher concentrations, second-nearest
neighbours having the peak of the distribution function at $r_{2}\approx
0.74n_{D}^{-1/3}$, and third--nearest neighbours ($r_{3}\approx
0.8n_{D}^{-1/3}$), also contribute into the interaction. It is easy to
estimate that at $10^{15}cm^{-3}\lesssim n_{D}\lesssim 10^{16}cm^{-3}$ the
interaction with the nearest neighbour dominates, though second and third
neighbours also contribute. Therefore, the correlation time can be estimated
as:

\begin{equation}
\tau _{c}\approx \hbar /\xi J(r_{c})  \label{4}
\end{equation}

where $r_{c}=\beta n_{D}^{-1/3},$ $\beta $ and $\xi $ are numerical factors
of the order of one, $J(R)=0.82E_{B}(R/a_{B})^{5/2}\exp (-2R/a_{B})$
(Ref.19). The value $r_{c}=\beta n_{D}^{-1/3}$ has the meaning of the
average characteristic distance between effectively interacting donors at
the given concentration. Therefore, one should expect $\beta $ to be in
between 0.54 and 0.8. Fig.3 shows that a good fit to the available
experimental data for bulk samples by the Eq.(4) is reached at $\beta =0.65$
, $\xi =0.8$ (Fig.3, dotted line). In spite of some scattering of
experimental points, the agreement with the model at very reasonable values
of parameters is remarkable. This is indeed an evidence that $\tau _{c}$ in
this concentration range is governed by the isotropic part of exchange
interaction. One cannot exclude, however, that there exist other physical
processes dominating the decay of the single-donor spin correlation (Eq.(3))
at low donor concentration, where the exchange interaction is ineffective.
Since experimental data in this concentration range are insufficient, we
consider it premature to include in the theoretical treatment specific
mechanisms of the correlation decay which may be relevant here (see
discussion at the end of the previous section). In the following, we will
use the experimentally determined values of $\tau _{c}$ to calculate spin
relaxation times.

With the knowledge of the concentration dependence of $\tau _{c}$, it
becomes possible to calculate the contributions into the spin relaxation
rate coming from hyperfine interaction and from anisotropic exchange
interaction, and therefore to find out $\tau _{s}$ in the insulating phase.
The expression for the spin relaxation time of donor-bound electrons due to
hyperfine interaction with lattice nuclei was derived by Dyakonov and Perel%
\cite{13}. At zero external magnetic field it reads:

\begin{equation}
\frac{1}{\tau _{SN}}=\frac{2}{3}\left\langle \omega _{N}^{2}\right\rangle
\tau _{c},  \label{5}
\end{equation}

where $\omega _{N}$ is the frequency of the electron-spin precession in an
effective fluctuating magnetic field produced by the nuclear spins within
the electron orbit. For shallow donors in GaAs $\left\langle \omega
_{N}^{2}\right\rangle ^{1/2}=2\cdot 10^{8}s^{-1}$ (Ref 10). The spin
dynamics of isolated localized electrons interacting only with nuclei (this
case is possibly realized at donor concentrations of the order of, or less
then, $10^{14}cm^{-3}$) has been considered theoretically in Refs.14 and 20.
Eq.(5), valid when $\left\langle \omega _{N}^{2}\right\rangle ^{1/2}\tau
_{c}\ll 1$, is a result of motional averaging of the random hyperfine
fields, acting upon the electron spin. As discussed above, the motional
narrowing at $n_{D}>1\cdot 10^{15}cm^{-3}$ is most likely due to rapid
flip-flop transitions induced by the exchange interaction. In the ensemble
of randomly distributed donors, these flip-flop transitions can be
interpreted as jumping of a chosen spin over different donors. The spin, on
the average, spends the time equal to $\tau _{c}$ at each of the donors it
visits. Due to the anisotropy of the exchange interaction, each jump is
accompanied by rotation of the spin through a small angle $\gamma $. This
results in spin relaxation with the characteristic time $\tau _{sa}$, given
by the expression:

\begin{equation}
\frac{1}{\tau _{sa}}=\frac{2}{3}\left\langle \gamma ^{2}\right\rangle \tau
_{c}^{-1},  \label{6}
\end{equation}

The mean squared value of $\gamma $ as a function of the inter-donor
distance $R$ can be calculated numerically using Eq.(16) of Ref.15, which
gives the following approximate expression for $\left\langle \gamma
^{2}(R)\right\rangle $ valid within the range of inter-donor distances from
1 to 20 Bohr radii :

\begin{eqnarray}
\left\langle \gamma ^{2}(R)\right\rangle &=&\frac{\alpha \hbar ^{3}}{m\sqrt{%
2mE_{g}}E_{B}a_{B}^{3}}\times  \label{7} \\
&&\times \left( 0.323+0.436\left( \frac{R}{a_{B}}\right) +0.014\left( \frac{R%
}{a_{B}}\right) ^{2}\right)  \nonumber
\end{eqnarray}

where $m$ is the electron mass, $E_{B}$ and $a_{B}$ are the Bohr energy and
the Bohr radius of the donor-bound electron, respectively; $\alpha $ is a
dimensionless factor at the cubic in $k$ term in the conduction-band
Hamiltonian (Chapter 2 of Ref.4). For GaAs, $\alpha $ is known to be about
0.07 (Chapter 3 of Ref.4); here we use the value 0.063, determined in Ref.21
from spin-flip Raman scattering.

We took $R_{av}=0.65\left( n_{D}\right) ^{-1/3}$ for the average inter-donor
distance relevant for the exchange interaction, as the above considerations
suggest. The solid line in Fig.3 represents the theoretical concentration
dependence of $\tau _{s}$, calculated as $\tau _{s}=\left( 1/\tau
_{sn}+1/\tau _{sa}\right) ^{-1}$. The concentration dependence of the
correlation time $\tau _{c}$ at $5\times 10^{14}cm^{-3}<n_{D}<4\times
10^{15}cm^{-3}$ is taken from the experiment, while an extrapolation by
Eq.(4) is used at $4\times 10^{15}cm^{-3}<n_{D}<2\times 10^{16}cm^{-3}$.

\subsection{Metallic phase $(n_{D}>2\times 10^{16})$}

The spin relaxation time at donor concentrations over 2$\times 10^{16}$ cm$%
^{-3}$ , i.e. in the metallic phase, has been calculated assuming that the
electron mean spin is accumulated near the Fermi level and that the Fermi
energy $E_{F}$ $\gg k_{B}T$. According to Chapter 3 of Ref.4, if the
electron momentum scattering is dominated by collisions with charged
impurities, the spin relaxation time of electrons with energy $E$ is:

\begin{equation}
\tau _{S}=\frac{315}{16}\alpha ^{-2}\frac{\hbar ^{2}E_{g}}{E^{3}\tau _{p}(E)}
\label{8}
\end{equation}

where $\tau _{p}$ is the momentum relaxation time. In the degenerate case we
deal with, $E$ stands for the Fermi energy $E_{F}=(3\pi ^{2})^{2/3}\hbar
^{2}n_{D}^{2/3}/2m$. To calculate $\tau _{p}$ as a function of $n_{D}$, we
used the Brooks-Herring method \cite{21}, i.e. evaluated, in the Born
approximation, the scattering cross-section of an electron off the Coulomb
potential screened by the degenerate electron gas. This approach gives the
following expression for $\tau _{p}$:

\begin{equation}
\frac{1}{\tau _{p}}=\frac{\pi n_{D}e^{4}}{\varepsilon ^{2}E_{F}^{3/2}\sqrt{2m%
}}\left[ \ln \left( 1+x\right) -\frac{x}{1+x}\right]  \label{9}
\end{equation}
where $x=\frac{8mE_{F}r_{0}^{2}}{\hbar ^{2}}=3^{1/3}\pi
^{5/3}a_{B}n_{D}^{1/3}$ , and the screening radius $r_{0}=\frac{1}{2}\left( 
\frac{\pi }{3}\right) ^{1/6}\left( a_{B}n_{D}^{-1/3}\right) ^{1/2}$.
Substituting Eq.$\left( 9\right) $ into Eq.$\left( 8\right) $, and assuming
that $E=E_{F}$, we obtain the formula for the spin relaxation time:

\begin{equation}
\tau _{S}=\frac{315}{16}\alpha ^{-2}\frac{E_{g}}{\pi ^{5}\hbar
^{3}a_{B}^{2}n_{D}^{2}}\left[ \ln \left( 1+x\right) -\frac{x}{1+x}\right]
\label{10}
\end{equation}

which was used to calculate the theoretical curve for $\tau _{s}\left(
n_{D}\right) $ at $n_{D}>2\cdot 10^{16}cm^{-3}$.

\bigskip

One can see that the theory demonstrates a fairy good agreement with the
experimental data all over the studied concentration range, both in
dielectric and in metallic phase. A slight systematical shift of the
calculated curve towards shorter $\tau _{s}$ in the metallic region may be
due to overestimation of the momentum relaxation time in our calculations.
Measurement of the low-temperature electron mobility along with the
experiments on spin orientation may be helpful in order to clarify this
point. And, of course, the peculiarity observed near MIT demands for
detailed experimental and theoretical studies.

\section{Conclusion}

Our results show that natural limits for the low-temperature spin lifetime
in bulk GaAs and other cubic compound semiconductors are placed by
stochastic precession of electron spins in random fields created by the
hyperfine interaction and by the spin-orbit interaction. Against commonplace
expectations, the crossover between these two main modes of spin decay in
GaAs occurs not at the metal-to-insulator transition ($n_{D}$ = 2$\cdot $10$%
^{16}$cm$^{-3}$) but at lower donor concentrations ($n_{D}\approx
(2-4).10^{15}$cm$^{-3}$), where electrons are bound to donors. A peculiarity
related to the metal-to-insulator transition (MIT) is clearly seen in the
dependence of the spin lifetime on doping near $n_{D}$ = 2$\cdot $10$^{16}$cm%
$^{-3}$. This peculiarity is due to changing the specific mechanism through
which the spin-orbit coupling affects the spin lifetime: in the metallic
phase it is the Dyakonov-Perel (DP) mechanism, while in the insulator phase
it is the anisotropic exchange interaction. Maximal value of $\tau _{s}$ of
free Fermi-edge electrons in heavily-doped samples is reached just above the
metal-to-insulator transition, where the Dyakonov-Perel relaxation is the
weakest. Another maximum of $\tau _{s}$ is in the dielectric phase, at an
optimal concentration determined by the interplay of the hyperfine
interaction and the anisotropic exchange interaction. Specifically in bulk
GaAs this is the absolute maximum of the spin lifetime, about 200 ns.
However, this value is the lifetime of the mean spin of the entire electron
ensemble. The spin lifetime at an individual donor, often discussed in
relation to quantum information processing, is limited either by the period
of precession in the fluctuation nuclear field ($\approx 5ns$), or by the
spin transfer to other donors, characterized by the correlation time $\tau
_{c}$. In our experiments, $\tau _{c}$ never exceeded $20ns$; in samples
with the longest spin relaxation times ($\tau _{s}\approx 180ns$ at $%
n_{D}\approx (2-4)\times 10^{15}cm^{-3}$), $\tau _{c}$ was of the order of
0.2ns. $\tau _{c}$ is a very important parameter that determines the
relative contributions of hyperfine and spin-orbit interactions and,
ultimately, the spin lifetime of localized electrons for a given
semiconductor. In bulk GaAs samples at $n_{D}>5\times 10^{14}cm^{-3}$, it is
governed by the exchange hopping of the electron spin over the impurity
band. At lower concentrations, it may be affected by other processes, for
instance, by exchange interaction with delocalized electrons\cite{10}. This
fact opens a possibility to realize optical or electrical control over the
spin lifetime of localized electrons in semiconductor structures.

\section{Acknowledgments}

We are grateful to I.A.Merkulov for very helpful discussions. Partial
support of the Russian Basic Research Foundation (grants 00-02-16991 and
01-02-17906), CRDF (grant RP1-2252), Volkswagen Foundation, UK EPSRC,
DARPA/SPINS, and ONR is acknowledged.

\begin{figure}[t]
\caption{Spectra of photoluminescence (PL) intensity (solid lines) and of
the PL circular polarization (dash lines) in GaAs: a) 0.1mkm thick GaAs
layer with electron concentration $n_{D}-n_{A}\approx 10^{14}cm^{-3}$
(insulating). Spectra taken in zero magnetic field under excitation by light
with the photon energy $h\nu =1.519eV$ and intensity $W=40mW/cm^{2}$ . b) 2$%
mkm$ thick GaAs layer with $n_{D}-n_{A}\approx 4.6\times 10^{16}cm^{-3}$
(metallic), spectra measured at $h\nu =1.520eV$ and $W=2W/cm^{2}$.}
\label{fig1}
\end{figure}

\begin{figure}[t]
\caption{Magnetic depolarization of photoluminescence (Hanle effect) at pump
densities $W=4W/cm^{2}$ (circles) and $W=0.5W/cm^{2}$ (squares).
Experimental values of the circular polarization degree $\rho _{c}$ are
divided by $\rho _{c}\left( B=0\right) $ . Solid lines: fit by Lorenzians
with half-widths of 8 G and 4 G. Inset: the Hanle-effect half-width as a
function of pump density. Extrapolation to zero pump gives $B_{1/2}=3.4G$,
corresponding to the spin relaxation time $\tau _{s}=76ns$.}
\label{fig2}
\end{figure}

\begin{figure}[t]
\caption{Spin relaxation time $\tau _{s}$ and spin correlation time $\tau
_{c}$ as functions of donor concentration in n-GaAs. Solid lines: theory.}
\label{fig3}
\end{figure}

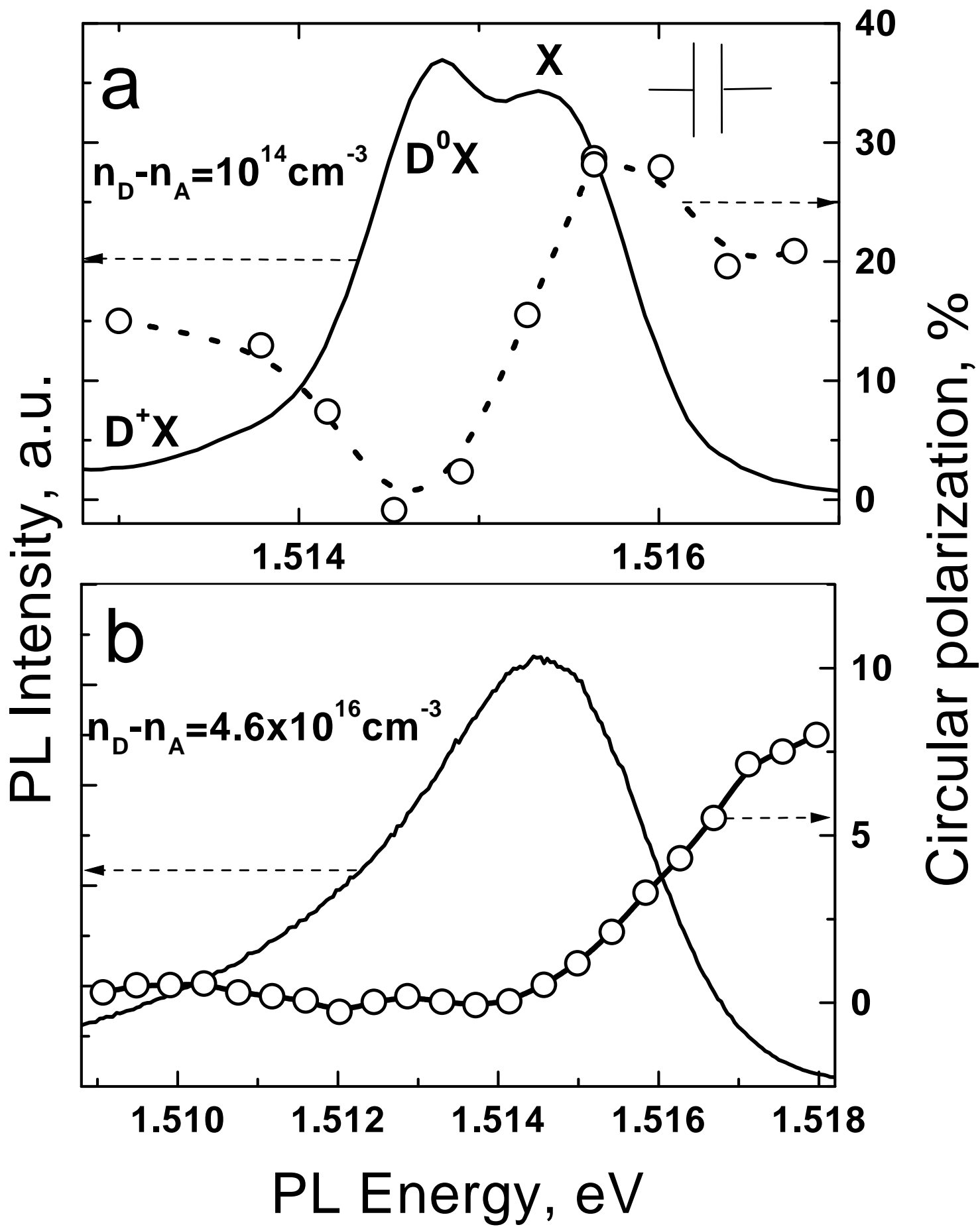
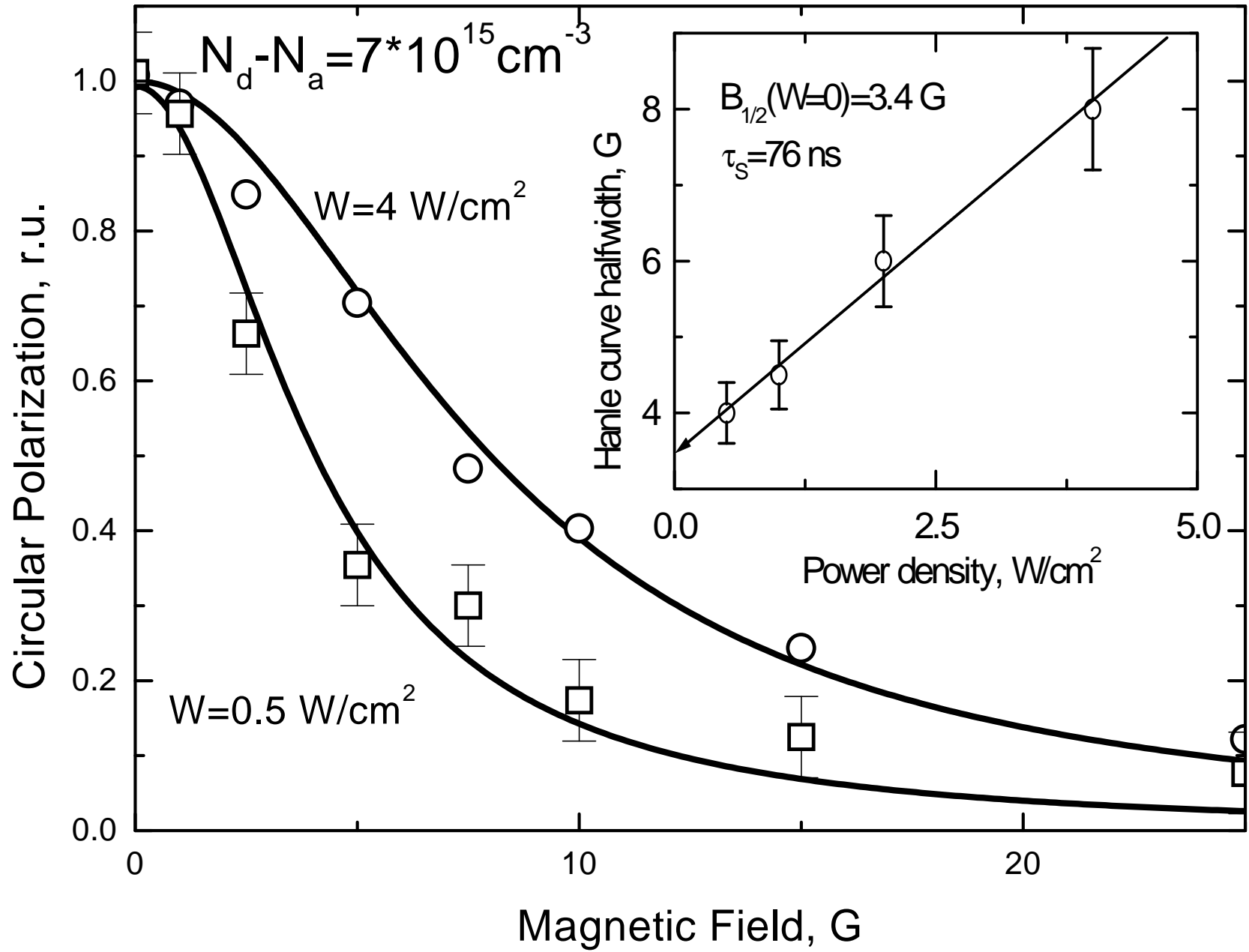
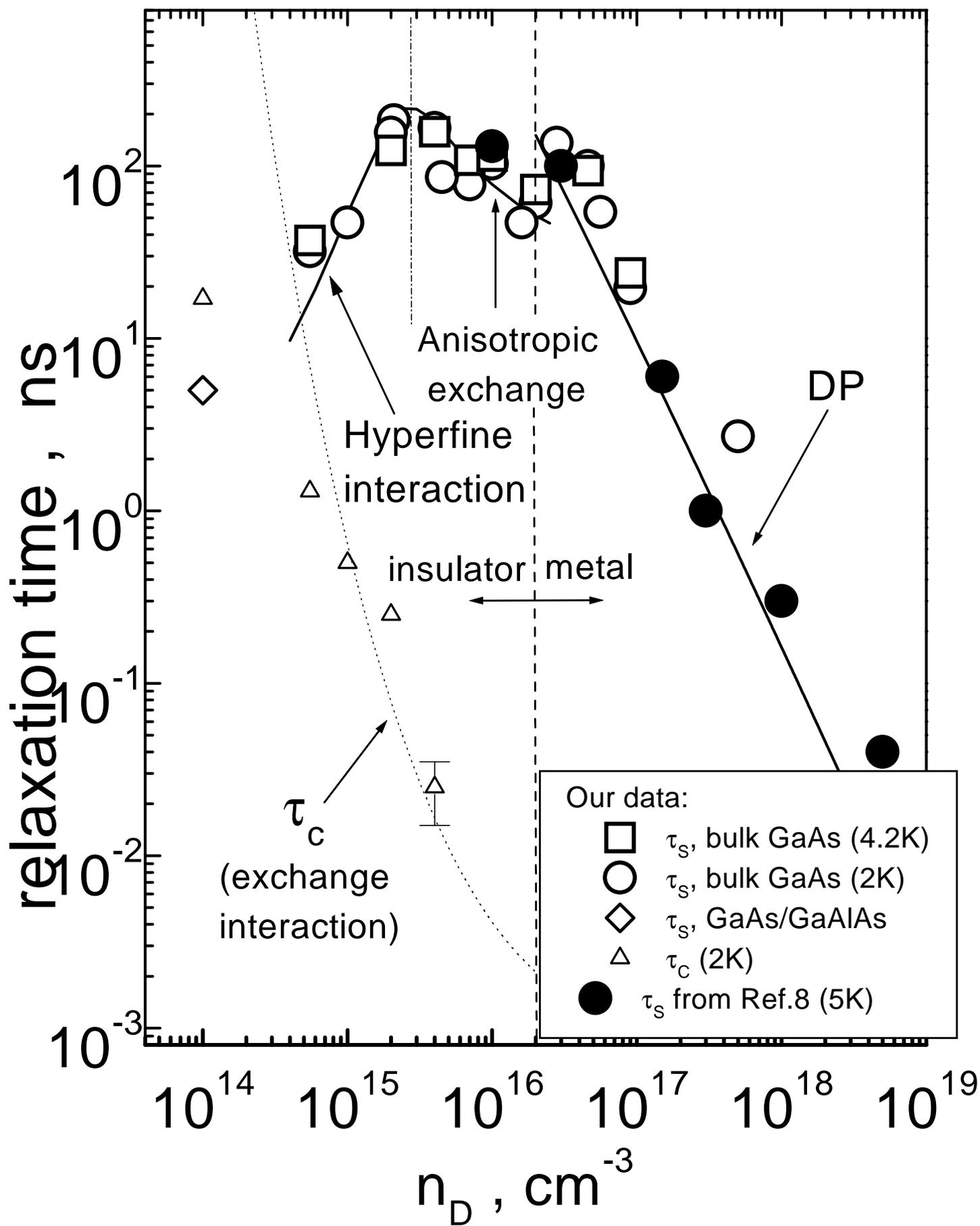
\end{document}